 \RequirePackage[2020-02-02]{latexrelease}
\documentclass[a4paper,floatfix,aps,twocolumn,groupedaddress,showpacs,english,superscriptaddress]{revtex4}
\usepackage[T1]{fontenc}
\usepackage{ragged2e}
\usepackage{appendix}
\usepackage{lscape}

\usepackage{wrapfig}
\usepackage{lscape}
\usepackage{rotating}

\usepackage[latin9]{inputenc}
\usepackage{float}
\usepackage{amstext}
\usepackage{graphicx}
\usepackage{esint}
\usepackage{color}
\usepackage[left=20mm,top=22mm,right=15mm,bottom=20mm]{geometry}

\makeatletter

\@ifundefined{showcaptionsetup}{}{%
 \PassOptionsToPackage{caption=false}{subfig}}
\usepackage{subfig}
\makeatother

\usepackage{babel}
\begin{document}
\title{Generating and processing optical waveforms using spectral singularities}
\author{Asaf Farhi}
\affiliation{Department of Applied Physics, Yale University, New Haven, Connecticut
06520, USA}

\author{Alexander Cerjan}
\affiliation{Center for Integrated Nanotechnologies, Sandia National Laboratories, Albuquerque, New Mexico 87185, USA}

\author{A. Douglas Stone}
\affiliation{Department of Applied Physics, Yale University, New Haven, Connecticut
06520, USA}
\affiliation{Yale Quantum Institute, Yale University, New Haven, Connecticut 06520,
USA}

\begin{abstract}
We show that a laser at threshold can be utilized to generate the class of coherent and transform-limited waveforms
$\left(vt-z\right)^{m}e^{i\left(kz-\omega t\right)}$ at optical frequencies.
We derive these properties analytically and demonstrate
them in semiclassical time-domain laser simulations. We then utilize these waveforms to expand other waveforms with high modulation frequencies and demonstrate theoretically the feasibility of complex-frequency coherent-absorption at optical frequencies, with efficient energy transduction and cavity loading. This approach has potential applications in quantum computing, photonic circuits, and  biomedicine. 
\end{abstract}

\maketitle
\section{Introduction}
There has been a recent explosion of interest in wave systems described by non-Hermitian operators, leading to complex eigenvalues in the general case. Such systems can exhibit singular behaviors not present in systems well described by a Hermitian operator.  In the current work we study the singularity associated with the onset of lasing as an optical waveform generator, and utilize it to generate waveforms associated with enhanced wave capture and absorption, phenomena which so far have been inaccessible at optical frequencies. 
We focus on singularities in the frequency domain Maxwell equations relating to the time-harmonic solutions of open electromagnetic systems, under asymptotic boundary conditions leading to solutions at discrete complex frequencies, in contrast to the conventional case of scattering boundary conditions (which lead to a continuous spectrum of real frequency solutions). 

The well-known example of this is the complex resonance spectrum of a scattering structure, where the boundary conditions at infinity are purely outgoing, without any input wave \cite{sauvan2013theory}. Another example is the complex spectrum arising from the boundary conditions of purely incoming waves at infinity; such boundary conditions can lead to the phenomenon of coherent perfect absorption, as we will define and discuss further below \cite{chong2010coherent,baranov2017coherent}. 
More recently, complex spectra associated with reflectionless scattering \cite{sweeney2020theory,dhia2018trapped} have been shown to have properties similar, but distinct from the two examples just cited.  All such spectra are defined by the boundary conditions on the wave operator at infinity, as just noted; or equivalently by the behavior of the linear scattering matrix (S-matrix) at the eigenfrequencies. At an eigenfrequency corresponding to a resonance, an S-matrix eigenvalue tends to infinity (a pole), whereas an eigenvalue of S tends to zero for the purely incoming case, and an eigenvalue for a sub-matrix of S tends to zero for reflectionless scattering.  Note that only when one of these eigenfrequencies is real is the corresponding state a {\it steady-state} solution of the linear Maxwell equation.


Singular behavior can arise in the case where two or more such eigenfrequencies become degenerate, which, in the absence of continuous symmetry, almost always involves tuning one or more parameters associated with the scattering structure (henceforth referred to as a resonator for convenience). 
 Generically such a degeneracy results in coalescence of the two eigenmodes as well, unlike the hermitian case. The point in parameter space where this happens is referred to as an exceptional point (EP); there is now an extensive theoretical and experimental literature on the physics of systems at an EP \cite{bender1998real,makris2008beam,moiseyev2011non,miri2019exceptional,ruter2010observation}. The case of resonant EPs is the most widely studied \cite{bender1998real,el2018non,guo2009observation}, but in the discussion of wave capture below we will focus on a different kind of EP. 

 The resonance spectrum has special properties not shared by the complex spectra associated with wave capture just mentioned.  Resonances of passive systems cannot occur on the real frequency axis; in active systems (lasers/amplifiers) resonances can reach the real axis (typically only singly).
This corresponds to the threshold for single-mode CW laser emission from the resonator, and the corresponding eigenfunction is the threshold lasing mode.
By definition this also corresponds to having a pole of the S-matrix on the real axis. 
Due to this pole the response to a small input will grow indefinitely in time, and we will exploit this property of the lasing singularity for waveform generation below \cite{mostafazadeh2011optical,bergman1980theory,farhi2016electromagnetic,farhi2020three,ge2010steady}. 

\subsection{Real and Virtual CPA}
The laser at threshold is related by time-reversal to a Coherent Perfect Absorber (CPA) \cite{chong2010coherent}. Quite generally, the Maxwell wave equation with outgoing boundary conditions for a finite resonator described by a susceptibility $\chi ({\bf r})$, maps under time reversal to the same wave equation applied to a resonator with susceptibility $\chi^* ({\bf r})$, and incoming boundary conditions. If the susceptibility is real, this implies that the eigenfrequencies corresponding to the zeros of the S-matrix are the complex conjugates of the resonance frequencies and must occur in the upper half complex plane, and not on the real axis. 
However if the resonator has gain and is at the lasing threshold, then this time-reversal mapping implies that the system with equivalent loss perfectly absorbs the time-reverse of the lasing mode, at the same frequency.  Such a resonator, tuned to have the correct amplitude and spatial distribution of absorption is called a Coherent Perfect Absorber, because it traps and then perfectly absorbs a specific coherent spatial waveform or wavefront (and only that input) \cite{chong2010coherent,baranov2017coherent}. 

CPA is the generalization to arbitrary multichannel scattering of the concept of critical coupling to a lossy resonator with a single input channel. However, it has recently been emphasized that zeros of the S-matrix {\it off} the real axis, which cannot be accessed with steady-state harmonic excitation, {\it can} be accessed transiently with a complex exponential input $A \exp[-i(\omega_z + i\gamma_z)t $, where $\omega_z + i\gamma_z$ is the eigenfrequency of the zero of the S-matrix \cite{baranov2017coherentvirtual}. Note that this waveform has an exponentially growing amplitude at rate $\gamma_z$ and can only be applied transiently. Accessing this "virtual" CPA does not correspond to actual absorption if the system is lossless, but rather to energy buildup and storage in the resonator until the exponential ramp is turned off.  It should be noted that such a concept of reversing decay from a resonator or even spontaneous emission from a single atom has also been suggested without connection to the CPA concept \cite{heugel2010analogy}.  Finally, there is a third important case \cite{ra2020virtual}, noted, but not studied in detail in earlier work.  If the resonator is not lossless, but doesn't have enough absorption to bring the zeros to the real axis, then each zero will have an imaginary frequency $\gamma_z \approx (\gamma_0 - \gamma_{a})$, where $\gamma_a$ is the absorption rate, and $\gamma_0$ is the imaginary part of the frequency of the lossless resonator. When such a resonator is excited with the appropriate complex frequency, no scattering will occur while the drive is on, but during this period the resonator will both accumulate energy and dissipate part of that energy via absorption, so that when the drive is turned off, some but not all of the incident energy will be released.  We will study such a case below. 

\subsection{Absorbing Exceptional Points: Frequency and Time Domain}

The existence of such wave capture processes, suggested that there would be interesting new behaviors when a resonator's parameters were tuned to a CPA EP, where two such eigenfrequencies become degenerate. For the case of real$-\omega$ CPA, this was explored in the frequency domain first by Sweeney et al.  \cite{sweeney2019perfectly}, where an anomalous lineshape for the absorption dip to zero reflection was predicted and later observed \cite{wang2021coherent}.  More recently, two of the authors and coworkers have extended the concept of virtual CPA to virtual CPA EP \cite{farhi2022excitation}, and studied the excitation of CPA EPs in the time domain.  
It was shown that the signature of the EP was the perfect absorption/capture of waveforms with the time-dependence $E(t) = B t\exp[-i(\omega_z - i\gamma_z)t]$ and $A \exp[-i(\omega_z + i\gamma_z)t] $ {\em in any coherent superposition}.  This statement also holds for the case $\gamma_z =0$,  and corresponds to the real CPA EP mentioned above; here the resonator absorbs a linearly growing plus constant wave envelope (i.e. without exponential growth) \cite{farhi2022excitation}.
For a one dimensional geometry, since the solution must satisfy the wave equation, the spacetime waveform which is perfectly absorbed at a real or virtual CPA EP is 
$E\propto  A \exp[-i(\omega_z + i\gamma_z)/v(z-vt)] + B\left(vt-z\right)  \exp[-i(\omega_z + i\gamma_z)/v(z-vt)]$.  This generalizes to higher order EPs (degeneracy $m>2$) where the relevant waveform is $\sum_{n=1}^m a_n \left(vt-z\right)^{m}  \exp[-i(\omega_z + i\gamma_z)/v(z-vt)].$ Interestingly, all the coherent waveforms within this class have
a frequency content at a single $\omega$ for infinite time pulses i.e., at $z=0$  $\mathcal{F}\left(t^{m}e^{i\omega_1t}\right)=\delta^{\left(m\right)}\left(\omega-\omega_{1}\right)$ \cite{farhi2022excitation}, which implies that for finite pulses they are transform limited.
While there will always be some imperfection in the wave capture due to the transient effects of the turn-on and turn-off of the waveform, it was demonstrated in that work that the high-order waveforms have substantially improved performance
in wave capturing. \cite{farhi2022excitation}.  Experimental realizations in microwave circuits confirm the basic concept \cite{mekawy2023observation}. 

Up to this point all the excitation properties we have discussed are independent of geometry or dimensionality. We have assumed that the appropriate input eigenvector of the S-matrix is imposed, which in general means exciting all the scattering channels with a specific coherent superposition or wavefront.  
For simplicity, here we will focus on the time-domain behavior in the simplest case of a one-dimensional single-sided slab resonator with a perfect mirror at the origin, so that the S-matrix reduces to a single scalar reflection amplitude coefficient, $r(\omega)$. However, the results we present will generalize to arbitrary geometry in a similar manner, by replacing $r(\omega)$ with the relevant eigenvalue of the S-matrix, $\sigma (\omega)$, and the plane wave input, $e^{ikz}$, with the relevant eigenfunction of the S-matrix. 

While it is possible (although not trivial) to generate transient exponentially rising waveforms at microwave modulation frequencies \cite{hinney2023efficient}, this isn't by any means straightforward to achieve at higher modulation frequencies, up to optical, making accessing virtual CPA quite difficult.  This suggests that using an all optical means of waveform generation might be necessary and useful. One type of optical processing was identified using a CPA in Ref. \cite{farhi2022excitation}. It was shown that a CPA converts the waveform $\left(vt-z\right)e^{i\left(kz-\omega t\right)}$ to a standard plane wave; in fact more generally a CPA can be used as a first-order envelope differentiator \cite{sol2022meta}. The conversion process just mentioned is the time reverse of the process we will study below.
Here we show that a standard continuous (CW) laser at threshold can act as a processor and potentially a generator for different time-varying waveforms (envelopes) at optical frequencies relevant to creating novel wave capture, enhanced absorption and reduced reflection. 
In Sec. II we present our results relating to waveform generation by a laser at threshold.  In the first subsection we analyze the linear response of the system in the time domain. In the second we include the critical effects of saturation for long times. In Sec. III we apply the new waveforms to wave capture (lossless case) and wave capture and enhanced absorption in the partially lossy case. In Sec. IV we summarize our results.



 
\section{Waveform generation by a laser at threshold}

The previous works reviewed above suggest that it would be interesting and potentially useful to generate the growing waveforms that can be efficiently absorbed at a real CPA EP, or to generate approximations to the exponentially rising waveforms captured at a virtual CPA (or virtual CPA EP). Such waveforms are not generated in an emission process. However, the ``native'' conversion process of a laser at threshold is to take a constant amplitude input wave and output the linearly rising harmonic wave envelope of interest (see Fig. 1).  
Specifically, from time  reversal of the CPA response \cite{farhi2022excitation}, it can be concluded that a laser at threshold within the linear response regime will convert $e^{i\left(kz-\omega t\right)}\rightarrow  \left(vt-z\right)e^{i\left(kz-\omega t\right)}.$  In principle, this process can be repeated to generate higher-order polynomial envelopes.
The setup is illustrated schematically in Fig. 1, where we assume a single-sided slab cavity as described above. These waveform conversion processes are intrinsically optical with no electronic manipulation necessary, and thus can be very fast. The conversion sets in after only very few roundtrips within the cavity  \cite{slavik2008photonic}. If one uses a microlaser for the conversion processes the  equilibration time is on the order of 10fs and one can obtain many optical cycles of the desired waveform as we will show. 

We note that an important earlier work has demonstrated that a laser near threshold can function as an envelope integrator \cite{slavik2008photonic}. This work was focused on demonstrating accurate all-optical integration of input pulses of different temporal shape. The demonstration was done with a fiber laser using electroptically modulated pulses down to $60 {\rm ps}$ in duration and the laser was operated below threshold to avoid saturation effects.  They did not report a saturation time in their experiments, and did not consider it in their mathematical model. 

Here we treat theoretically a laser exactly at threshold (with the ideal linear transfer function prior to saturation) to generate waveforms with \emph{optical modulation frequencies}. We calculate the saturation time to determine the upper time limit on linear functioning.  
We consider an approximately square (constant amplitude) input pulse, which can be processed to generate the desired waveform.  We will discuss ways to generate such pulses after presenting the results assuming such pulses are available.

\subsection{Laser at Threshold: Linear response}

For the 1D slab resonator of Fig. 1 the S-matrix in the frequency domain is simplified to the reflection coefficient $r(\omega)$, which has the analytic form:
\begin{equation}
r(\omega)=-\frac{r_{1}+e^{2ikn_{1}l_{1}}}{1+r_{1}e^{2ikn_{1}l_{1}}}\approx-\frac{r_{1}+e^{2i\omega/cn_{1}l_{1}}}{2in_{1}l_{1}\left(\omega-\omega_{1}\right)/c}.
\end{equation}
Here $r_1$ is the reflection amplitude of the front mirror, $k=\omega/c$, $l_1$ is the cavity length, $n_1$  is a uniform index of refraction with uniformly distributed gain, to be described in more detail below. In the second approximate equality we have expanded $r (\omega)$ near a specific resonance frequency, $\omega_1$ (assumed to be at threshold). 
To calculate the response to transient inputs we need to include the turn-on (and turn-off) of the input.
In order to gain some intuition, we first express in the frequency domain the
incoming field at $z=0$ $E_{\mathrm{inc}}=\theta(t)e^{i\omega_1 t},$ where $\theta(t)$ is a step function, and the laser response $r (\omega)$ and the output in the vicinity of a resonance
\[
\mathcal{F}\left(E_{\mathrm{in}}\right)=\frac{1}{\omega-\omega_{1}},\,\,\,r\approx \frac{\mathrm{const}}{\omega-\omega_{1}},\,\,\,\mathcal{F}\left(E_{\mathrm{scat}}\right)\approx\frac{\mathrm{const}}{\left(\omega-\omega_{1}\right)^{2}}.
\]
where $\mathcal{F}$ is the Fourier transform.
Hence, taking into account causality, we expect to have in the time domain
\[
E_{\mathrm{scat}}\left(t\right)=\mathcal{F}^{-1}\left(E_{\mathrm{scat}}(\omega)\right)\approx te^{i\omega_1 t}\theta\left(t\right)
\]
In order to satisfy the wave equation, the electric field has to be
of the form $f\left(vt-z\right)$ and we obtain $E_{\mathrm{scat}}\left(t,z\right)\approx\left(vt-z\right)e^{i\left(kz-\omega t\right)}\theta\left(t\right),$ see
Fig. 1. 

Note that $r(\omega)$ has additional poles that will be off the real axis and some distance away
from the lasing frequency.  However, one has to take into account these poles
in the calculation. 
\begin{figure}[htp]
\includegraphics[width=8cm]{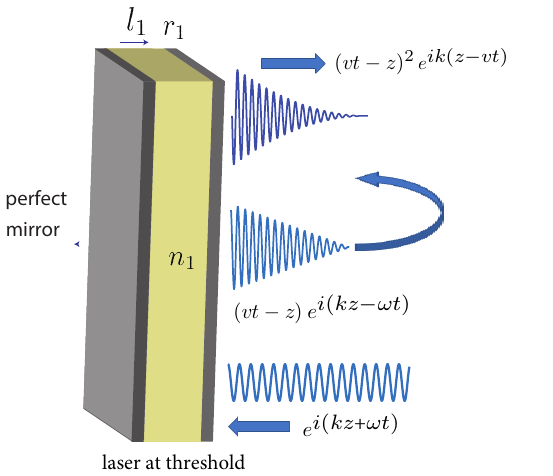}
\caption{A scheme of the setup: a plane wave impinging on a laser at threshold
that generates the waveform $\left(vt-z\right)e^{i\left(kz-\omega t\right)}$,
which in turn is converted to the waveform $\left(vt-z\right)^{2}e^{i\left(kz-\omega t\right)}.$}
\end{figure}
To that end, we write the inverse Fourier transform (IFT) of the output
explicitly. We analyze the IFT integral close to the divergence
at $\omega_{1}$ and in the other regions separately, similarly to
the calculation of the conversion process at a CPA \cite{farhi2022excitation}
\begin{eqnarray}
&E_{\mathrm{scat}}\left(t\right)=
\intop_{-\infty}^{\omega_{1}-\Delta\omega}\frac{r(\omega)}{\left(\omega-\omega_{1}\right)}e^{i\omega t}d\omega \nonumber\\
&+\intop_{\omega_{1}-\Delta\omega}^{\omega_{1}+\Delta\omega}\frac{a(\omega)}{\left(\omega-\omega_{1}\right)^{2}}e^{i\omega t}d\omega
+\intop_{\omega_{1}+\Delta\omega}^{\infty}\frac{r(\omega)}{\left(\omega-\omega_{1}\right)}e^{i\omega t}d\omega,
\end{eqnarray}
where $a(\omega)$ is the numerator of $r(\omega)$ and has only exponentials that will generate time shifts in the time domain. The contribution of the integral with the pole is expected to be dominant at large $t$ and it can be shown that at the large $t$ or large $\Delta\omega$
limit it agrees with the expression obtained by the simpler approximation for $E_{\mathrm{scat}}$ above, see Appendix A for details. From similar arguments one can show that when the waveform
$\left(vt-z\right)e^{i\left(kz-\omega t\right)}$ impinges on a laser
at threshold, it will be converted to $\frac{1}{2}\left(vt-z\right)^{2} e^{i\left(kz-\omega t\right)}.$
Thus, by inputting the output back to the laser or to another identical
laser, it is possible to generate any waveform of the class $\left(vt-z\right)^{m}e^{i\left(kz-\omega t\right)}.$ For a schematic of the setup see Appendix B.
 
From this inverse Fourier transform we can obtain a good approximation to the scattered field in response to an input square pulse starting at $t=0$ and ending at time $t_1$, assuming that saturation hasn't set in. The result is, 
\begin{eqnarray}
&\frac{E_{\mathrm{sc}}(t)}{E_{\mathrm{inc}}}=ie^{-i\omega_{1}t}
\frac{\left[t\text{sgn}(t)+\left(t_{1}-t\right)\text{sgn}\left(t-t_{1}\right)\right]*\left(1+r_{1}\delta\left(t-\frac{2l_{1}n_{1}}{c}\right)\right)}{\tau_{\mathrm{min}}}\nonumber\\
&\approx-\frac{ie^{-i\omega_{1}\tilde{t}}\left[\tilde{t}\text{sgn}(\tilde{t})+\left(t_{1}-\tilde{t}\right)\text{sgn}\left(\tilde{t}-t_{1}\right)\right]\left(1+r_{1}\right)}{\tau_{\mathrm{min}}},\,\tilde{t}=t-\tau_\mathrm{min}
\end{eqnarray}
where $*$ denotes  convolution, and $ \tau_\mathrm{min} = \frac{2l_{1}n_{1}}{c}$, is the round-trip time of light travel in the cavity.  For $\Delta\omega_{\mathrm{poles}}\gg\Delta\omega_{\mathrm{input}},$ only
the numerator determines the equilibration time to the linear envelope behavior, which is on the order of $\tau_\mathrm{min}$.  
From this expression the envelope in the linear regime has the form
\begin{equation}
|E_{\mathrm{sc}}(t)| \approx \frac{|E_{\mathrm{inc}}|(1 + r_1)}{\tau_{\mathrm{min}}}t.  
\end{equation}
In principle $n_1$ in the definition of $\tau_{\mathrm{min}}$ involves the gain medium and is complex but the gain term is a small correction, and is negligible in the unsaturated linear regime.

\subsection{Laser at threshold: Saturated response}

In order to calculate the full nonlinear response of the laser we will need to solve the semiclassical laser equations with nonlinear coupling between the gain medium and the wave equation, to demonstrate the conversion $e^{i\left(kz-\omega t\right)}\rightarrow \left(vt-z\right)e^{i\left(kz-\omega t\right)}$ over a finite time until saturation sets in.  To do this we performed finite-difference time domain (FDTD) semi-classical simulations in MEEP utilizing a recently developed laser module \cite{cerjan2020modeling,oskooi2010meep,johnson2021notes,taflove2005computational} with a an effective two level description, which approximately maps to a three-level system when there is a fast $3\rightarrow 2$ transition.  We choose the units so that the length of our cavity is $l_1= 4 \mu{\rm m},$ which is realizable experimentally \cite{hill2014advances}.  This has the effect of making the roundtrip time of $4\cdot 10^{-14}$s; the corresponding lasing frequency is in the red optical spectrum. We assume a smooth but few-cycle turn-on of the drive and run the simulation until saturation, whereas in the applications we envision the drive pulse would be turned off prior to saturation. In our FDTD simulations the cavity has uniformly distributed gain and is uniformly pumped. Details and further parameters for the laser simulations are given in Appendix C.

It is worth noting that in the simulations we neglect the effect of spontaneous emission, which results in noise in the output. To include such an effect in a simulation valid for both the linear and nonlinear response regimes is beyond the scope of the current work. In the linear response regime (constant gain, unsaturated inversion) one can approximate the spontaneous emission using Fermi Golden Rule and the resulting expressions
$ \Gamma=\frac{\pi\omega p^2}{\hbar \epsilon_0}\rho_\mu,  \,\,\, \rho=-\frac{2\omega}{\pi}\mathrm{Im}({G_{\mu \mu }(\mathbf{r},\mathbf{r})}).$
where $\rho_\mu$ is the density of electromagnetic states, $p$ is the dipole moment, and $G_{\mu \mu }(\mathbf{r},\mathbf{r})$ is Green's tensor \cite{carminati2015electromagnetic}. To calculate the Green tensor a full electrodynamic approach in three dimensions needs to be utilized since $l_1\gtrsim\lambda$ and the atoms should be modeled as a point source. We outline such an approach in Appendix D. The experiments using a fiber laser at threshold to integrate pulse envelopes found that the laser noise did not significantly degrade the output \cite{slavik2008photonic}. In addition, in practice the waveforms will be trimmed before saturation sets in; we neglect this effect, since we focus on the linear response-regime and the switching off shape will depend on the trimming mechanism.

Consistent with our analytic analysis, as shown in Fig 2a, the outgoing field (blue, incoming red) is indeed rising with a linear envelope. 
In Appendix E we plot the laser response for smaller and larger incoming field amplitudes,
which shows that the conversion process is robust to changes in the
amplitude of the incoming field and at low incoming field amplitudes
the envelope is wavy after the linear-response time window. We also derive in Appendix F the rate-equation analysis with a constant drive, which explains these dynamics. 
In Fig. 2b we show the behavior on the longer time scale, where saturation sets in.  First the field overshoots and then, without significant relaxation oscillations, relaxes to a lower steady-state output of constant amplitude, $E_{\mathrm{scat}}\propto E_{\mathrm{sat}} e^{i\omega t}.$
In Fig. 2c we present the scattered field as a function of $z$, both within
the laser cavity and in free space. The field in free
space has the form $E_{\mathrm{scat}}\propto\left(vt-z\right)e^{i\left(kz-\omega t\right)}$
and matches well the predicted spatial field distribution in Fig. 1.

The saturation field amplitude, $E_{\mathrm{sat}}$ cannot be calculated analytically, but since it does correspond to a steady-state it can be calculated independently using using Steady-state Ab Initio Laser Theory 
(SALT) \cite{ge2010steady,cerjan2011steady}, and specifically by the method of \cite{cerjan2014steady}, referred to as I-SALT, which solves the saturated wave equation with an injected signal.  In this approach one reduces the coupled Maxwell-Bloch laser equations to a non-linear frequency domain Maxwell wave equation with saturation by assuming a single-frequency constant amplitude harmonic response as we have here for long times.
The equation is solved self-consistently including spatial hole-burning effects as well as saturation, using a convenient basis set, providing an exact (up to numerical error) solution for $E_{\mathrm{sat}}$.  In Fig. 2d we show that the saturation field calculated by the two independent methods (FDTD and I-SALT) agree extremely well.

\begin{figure*}[htp]
\subfloat[]{\includegraphics[width=5.5cm]{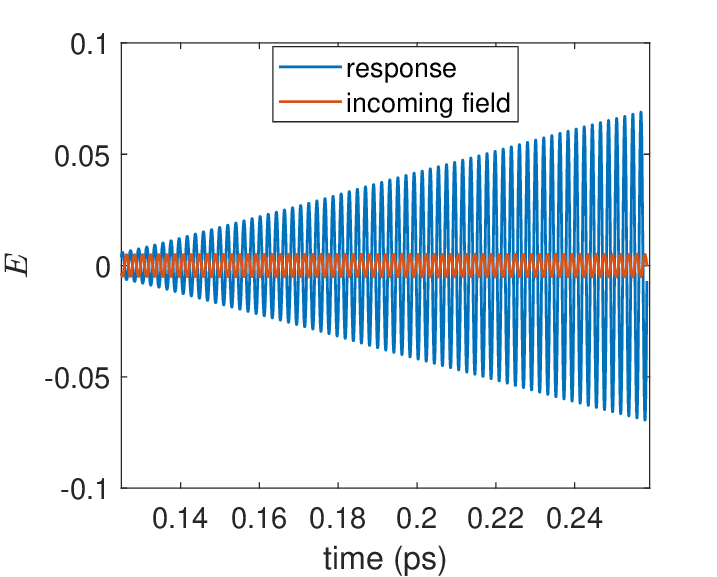}

}\subfloat[]{\includegraphics[width=5.5cm]{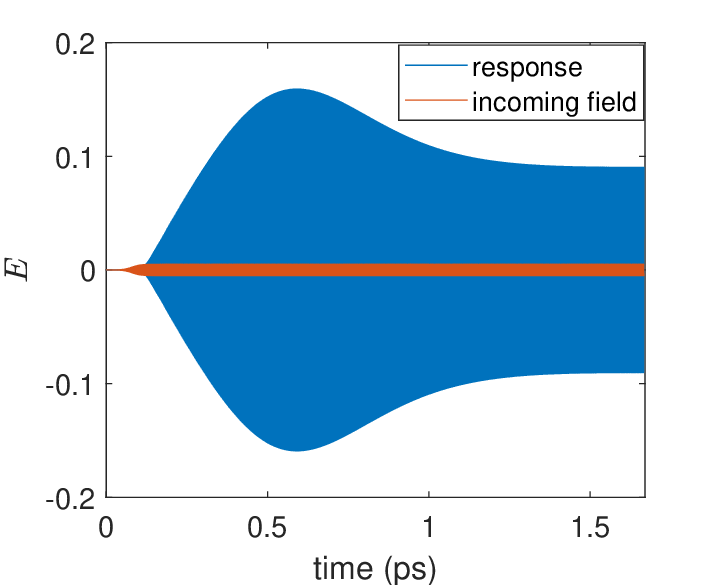}

}\subfloat[]{\includegraphics[width=5.5cm]{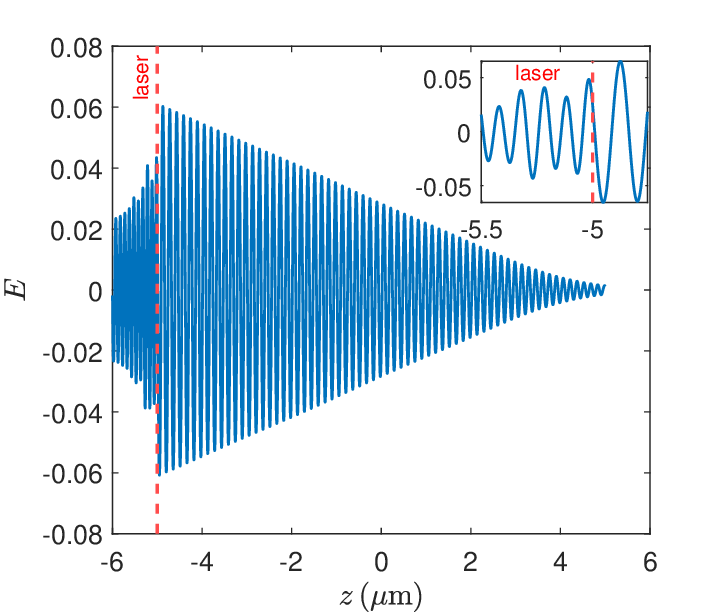}

}


\subfloat[]{{\includegraphics[width=5.5cm]{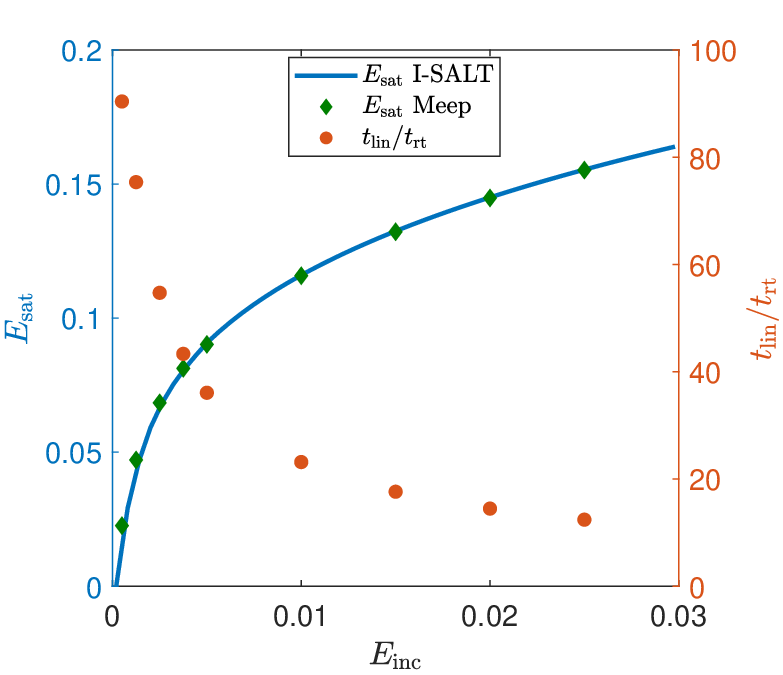}}

}\subfloat[]{
\includegraphics[width=5.5cm]{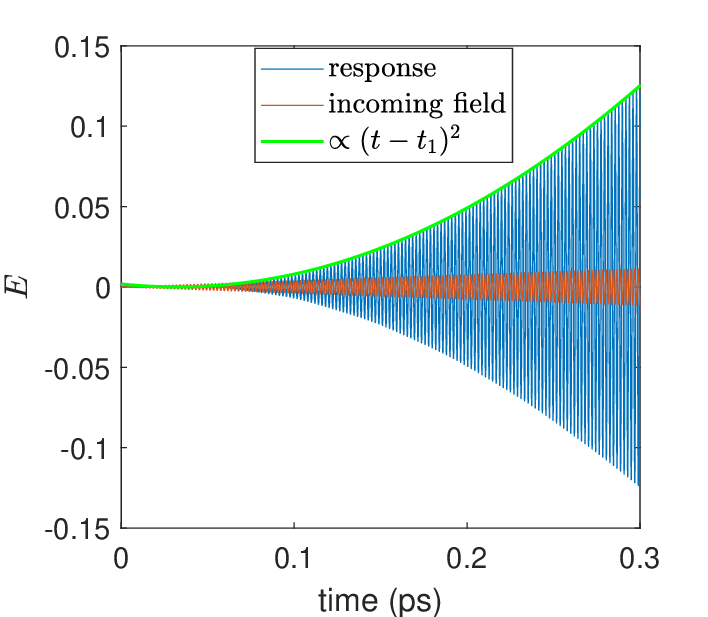}

}\subfloat[]{
\includegraphics[width=5.5cm]{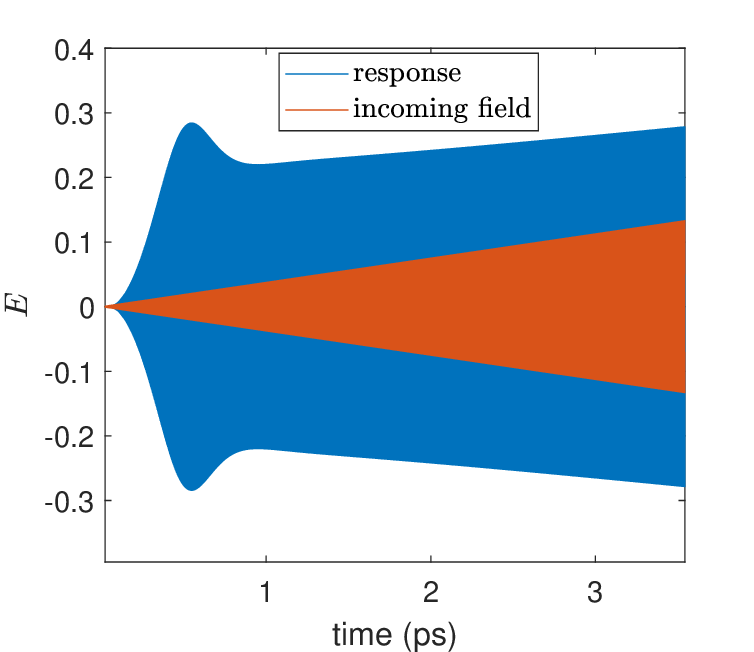}

}

\caption{\RaggedRight Semiclassical FDTD MEEP simulation results of the incoming and scattered electric fields with $J=0.01,\,\,E_\mathrm{inc}=0.005$ as functions
of time for 
(a) The entire simulation time and (b) the linear-response time window, and space (c), which confirm the output waveform $\left(vt-z\right)e^{i\left(kz-\omega t\right)}$
in Fig. 1. Note that in (c) the laser cavity is situated to the left of the vertical line and the gain medium is uniformly distributed along the cavity.  The saturated field and $t_\mathrm{lin}/t_\mathrm{rt}$ as functions
 of $E_{\mathrm{inc}}$ (d). Response of the laser to $E_{\mathrm{inc}}\propto\left(vt-z\right)e^{i\left(kz-\omega t\right)}$
 resulting in a scattered field of the form $\left(vt-z\right)^{2}e^{i\left(kz-\omega t\right)}$ in the linear regime (e) and (f).}
\end{figure*}


From knowledge of the saturation field we can place a lower bound on the time for which the linear response expressions will be valid.  We see from the simulations that the field is always linear at least until the level of $E_{\mathrm{sat}}$; typically it overshoots, becomes nonlinear and then relaxes. Hence the first time when $|E_{\mathrm{sc}}(t)| = E_{\mathrm{sat}} $ provides a lower bound on the saturation time, $t_{\mathrm{sat}}$.  From Eq. (4) we see that 
\begin{equation}
t_{\mathrm{sat}}  > \frac{|E_{\mathrm{sat}}| \tau_\mathrm{min}}{ E_{\mathrm{inc}}(1 + r_1)}.   
\end{equation}
This confirms that the system is linear for a time which scales as the ratio of the saturation field to the incident field, and can be quite long compared to the roundtrip time. We plot this lower bound to the saturation time in Fig. 2d.  Note that the roundtrip time is roughly ten times the optical period, so for the lowest input power we have used the system is linear for more than $\sim 10^3$ oscillations.  
In applications, by changing the cavity length or lowering the drive power one can tune the linear regime to exceed the pulse length of interest.





 \begin{figure*}

 \subfloat[]{\includegraphics[width=5.5cm]{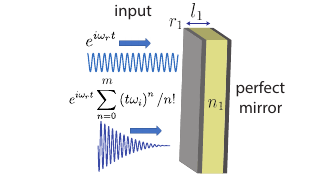}

  }\subfloat[]{\includegraphics[width=5.5cm]{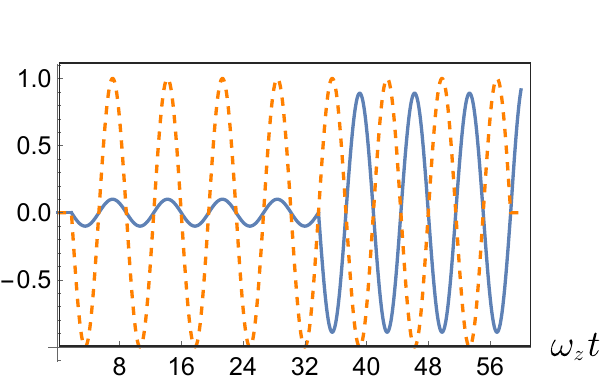}

  } \subfloat[]{\includegraphics[width=5.5cm]{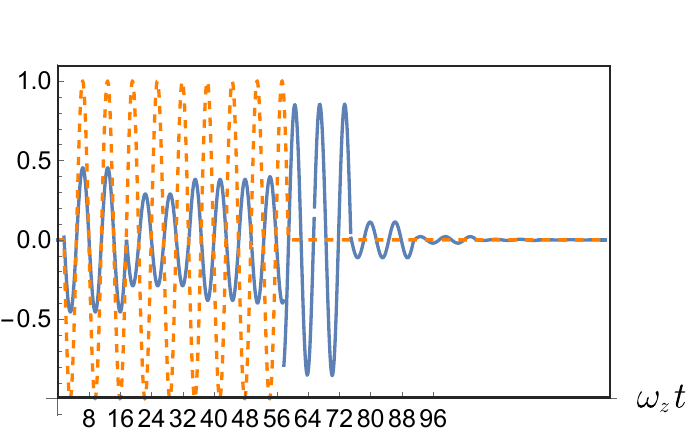}
}

 \subfloat[]{\includegraphics[width=5.5cm]{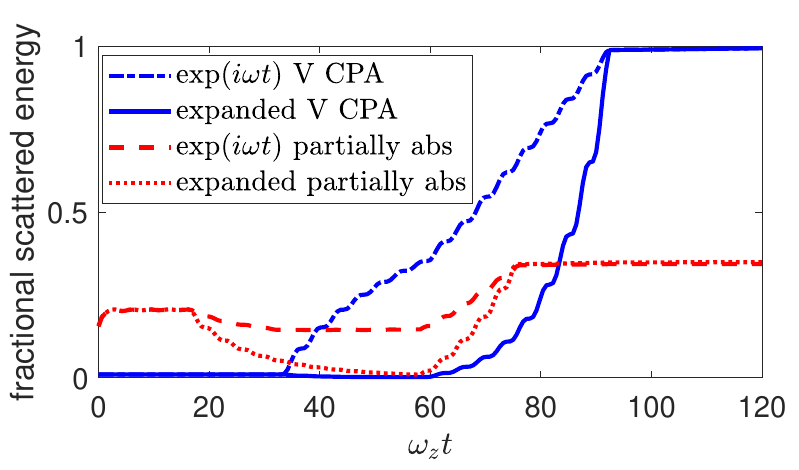}

 } \subfloat[]{\includegraphics[width=5.5cm]{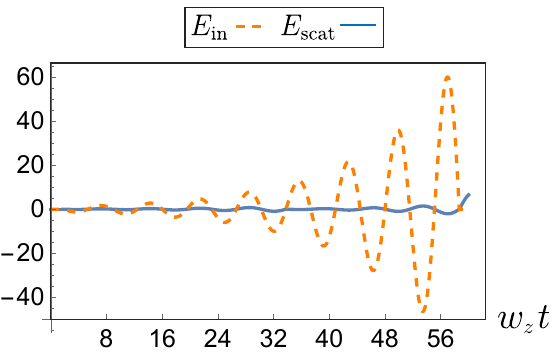}

  } \subfloat[]{\includegraphics[width=5.5cm]{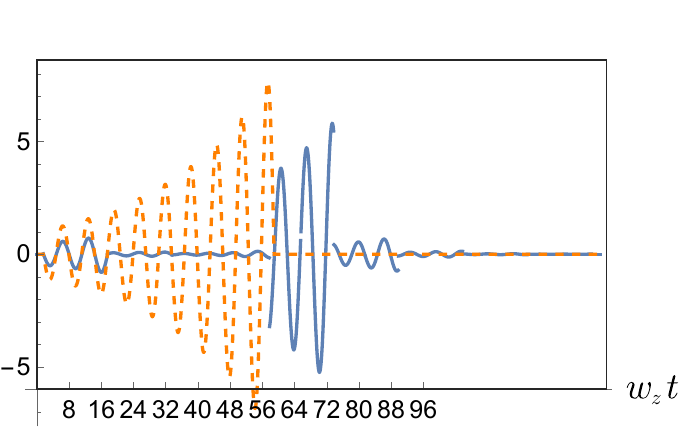}
 }

 \caption{
\RaggedRight
 Scattered fields for cavities with complex absorption eigenfrequency $\omega_z + i \gamma_z$; we contrast cases of constant amplitude illumination with that of an  optimized rising waveform (shown schematically in (a)). Figs. (b) and (e) are for illumination of a lossless cavity with $l_1n_1=4,\,\,n_1=1.22\,\,,\omega_z=3.53+0.287i$. The rising waveform (e) is much more efficient in capturing and storing energy than the constant amplitude input (b). Figs. (c),(f) are for scattering from a partially lossy, undercoupled cavity (zero above the real axis) with $l_1\mathrm{Re}(n_1)=4,\,\,n_1=1.22+0.072i,\,\,\omega_z=3.57+0.071i$.  Again, the optimized rising waveform (f) is much more efficiently captured during the drive than is the constant amplitude wave (c). Fig. (d) compares the fractional scattered energy $\intop_{0}^{t}E_{\mathrm{sc}}^{2}dt'/\intop_{0}^{t}E_{\mathrm{in}}^{2}dt'$ vs. $t$ for both resonators and both inputs with 178.7 and 16.3 times less fractional scattered energy before the switch off for the expanded wave for the lossless and lossy resonators, respectively.}
 \end{figure*}

In Fig. 2e we present the response of the laser to an incoming
field of the form $E_{\mathrm{inc}}\propto\left(vt-z\right)e^{i\left(kz-\omega t\right)}.$
As expected, in the linear-response time window, there is a conversion
of the input to $\left(vt-z\right)^{2}e^{i\left(kz-\omega t\right)}$, in agreement with our previous analysis. 
In Fig. 2f we present the laser response to this input for later times.  As before, the linear response overshoots and relaxes, but, distinct from the previous case, here it does not relax to a constant amplitude after saturation sets in, because the drive amplitude itself is increasing linearly.  In practice the length of this drive would be limited by the length of the input drive pulse, which we have simply neglected here.  Again, we do not envision using the laser converter in the saturated regime for the applications we discuss here.

In order to switch on and off these optical signals, one can use optical shutters, which are usually based on two-photon absorption \cite{dawes2005all}. Optical shutters were shown to exhibit 10fs switch off time \cite{yavuz2006all} and operate at low field intensities \cite{venkataraman2011few}. Thus, one could use a pulsed laser to switch on or off the optical signals that are emitted from the cw laser. 

\section{Wave capture: enhanced internal field and absorption}

The applications we have in mind (at least initially) are the capture, storage and absorption of optical energy, utilizing the incoming spectral singularities at complex and real frequencies.  The most natural example is to use our laser converter to take a roughly square pulse and generate a linearly rising pulse $(vt-z)e^{i(\omega t -kz)}$ with an oscillation frequency $\omega$ equal to that of a cavity tuned to have CPA EP. 
Inputting this waveform instead of $e^{i(\omega t -kz)}$ significantly reduces the energy lost to transient scattering when a cavity is tuned to a CPA EP, see Ref. \cite{farhi2022excitation} Fig. 3 b. While both waveforms are perfectly absorbed by such a cavity in steady-state, for a finite pulse 
the interaction of light with the cavity turns on more adiabatically when the linear ramp is applied, because the incident field is small while the system equilibrates to quasi-steady state and large when the system is perfectly absorbing. 

Here we focus on accessing just a single complex zero with a rising exponential by expanding it with the generated waveforms. This approach can also be applied to the second-order mode excitation of a virtual CPA EP.
We will assume that the absorbing cavity response always remains linear here, for simplicity.
If we excite the undercoupled cavity with the correct complex frequency rising waveform corresponding to a zero, this input will be perfectly captured in steady-state. 
The field intensity will continue to grow exponentially with a rate given by the imaginary part of the frequency. In this case the absorption per unit time will be enhanced, because of the constantly growing intensity in the cavity.
This effect can be used to enhance flux into an absorbing detector during a finite time interval.  
Please note that due to the exponential growth of the input, much of the total signal is contained in the last few roundtrip times of the pulse, which will not be trapped once the drive is turned off. Hence we will find that the total fractional absorption is not necessarily much higher, but the absorption during the equilibrated portion of the pulse should be greatly enhanced. Also, the field within the cavity should be strongly enhanced (this effect is largest for a lossless cavity), which can be useful for applications involving high-field or non-linear effects.
This reasoning shows the potential utility for absorption and wave capture of being able to generate rising exponential waveforms efficiently, something which is quite difficult to do at optical frequencies. 

We explore this approach in the calculations shown in Fig. 3, where we assume we can generate polynomials up to $m=4$ and utilize them for absorption/capture in a lossless and partially lossy, undercoupled cavity. 
Comparing Figs. 3b to 3e (lossless cavity) and 3c to 3f (partially lossy cavity) shows  that the expanded waveforms exhibit dramatically weaker scattering compared with the plane waves in both cases, once the transient loading period is over.  Fig  3d shows that before the input switches off, the fractional scattered energies of the expanded waveforms for the lossless and lossy cavities decrease to 0.2 and 0.9 percent, respectively, while for the constant amplitude input it remains at 35 and 14.5 percent, corresponding to much weaker relative scattering for the first case by factors of 178.7 and 16.3, respectively. For a discussion on the range of possible modulation frequencies using our approach see Appendix G.

\section{Conclusion}

In summary, we have shown that a laser at threshold can in principle be used to generate 
a class of waveforms at optical frequencies $\left(vt-z\right)^{m}e^{i\left(kz-\omega t\right)}$ by iterative processing of a constant amplitude input pulse (or by sending such a pulse through a laser array).  These waveforms can be used to excite a resonator or detector with an optimal rising waveform, which strongly enhances wave capture, and absorption if the cavity is lossy.  Effectively, this provides a way to critically couple to an undercoupled cavity during the equilibrated portion of the excitation.  For more general resonators with multiple input channel the appropriate coherent multichannel input must be applied. 
Exponentially rising waveforms 
 are well known to be of interest for cavity loading, with potential applications in quantum information processing \cite{wenner2014catching}, photonic circuits \cite{rios2015integrated}, and atom loading \cite{heugel2010analogy}. The increased field and absorption rate could be used for enhanced signal detection and light-matter interactions.

An interesting possibility would be to generate a second-order integrator via a laser with an exceptional point at threshold \cite{benzaouia2022nonlinear}. While the behavior in linear response of such an EP laser is easily shown to provide such an integrator, to implement an EP laser additional conditions would need to be satisfied \cite{benzaouia2022nonlinear}. We note that to successfully implement our setup a fast switch on of the pulse is required, with recent progress in this field \cite{dawes2005all,yavuz2006all,venkataraman2011few}.
The potential dramatic enhancement in temporal resolution and wave capture at optical frequencies provided by such waveforms seems to us novel and worthy of further study.  


\section*{Acknowledgments}
A.F. and A.D.S. acknowledge support from the Simons Foundations under the Collaboration on Extreme Wave Phenomena.
M. Yessenov, H. Suchowski, T. Schwartz, H. Diamandi, Y. Warshavsky, D. Hershkovitz, and A. Levanon are acknowledged for the useful comments. A.C. acknowledges support from the Laboratory Directed Research and Development program at Sandia National Laboratories. This work was performed, in part, at the Center for Integrated Nanotechnologies, an Office of Science User Facility operated for the U.S.\ Department of Energy (DOE) Office of Science. Sandia National Laboratories is a multimission laboratory managed and operated by National Technology \& Engineering Solutions of Sandia, LLC, a wholly owned subsidiary of Honeywell International, Inc., for the U.S.\ DOE's National Nuclear Security Administration under contract DE-NA-0003525. The views expressed in the article do not necessarily represent the views of the U.S.\ DOE or the United States Government.

\bibliographystyle{unsrt}

\newpage

\appendix
\begin{widetext}


 \section{The contribution of the second integral}
 Imposing causality the second integral in Eq. (1) in the main text reads
\begin{eqnarray}
&a(t)*\intop_{-\infty}^{\infty}\mathrm{window}\left(\frac{\omega-\omega_{1}}{\Delta\omega}\right)\frac{1}{\left(\omega-\omega_{1}\right)^{2}}e^{i\omega t}d\omega=\nonumber\\
&a(t)*e^{i\omega_{1}t}\mathrm{sinc}\left(t\Delta\omega\right)*te^{i\omega_{1}t}\theta\left(t\right)=
\frac{a(t)}{2\pi}*e^{i\omega_{1}t}\times\nonumber\\
&\left(2t\text{Si}(\Delta\omega t)+\pi t+2\cos(t)/\Delta\omega\right)\theta\left(t\right)\rightarrow a(t)*e^{i\omega_{1}t} t\nonumber
\end{eqnarray}
\noindent
where $\text{Si}(z)=\intop_{0}^{z}\sin\left(t\right)/tdt $ and $*$ denotes convolution.
From the single pole approximation we get
\begin{equation}
\frac{E_{\mathrm{sc}}(t)}{E_{\mathrm{inc}}}=-\frac{c}{4l_{1}n_{1}}\left[r_{1}+\delta\left(t-\frac{2l_{1}n_{1}}{c}\right)*\right]e^{-i\omega_{1}t}t\theta\left(t\right)
\end{equation}
\section{Suggested Implementation}

\begin{figure}[htp]

\includegraphics[width=15cm]{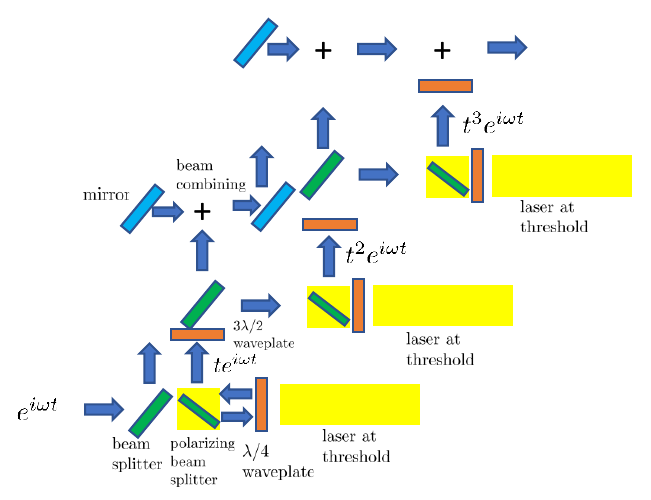}

\caption{A scheme of a suggested implementation. The design of the desired signal can be bottom to top. We start from the coefficient of the last Taylor expansion component, calculate the input in the last laser using the transfer function, and add to this input the coefficient of the second to last Taylor expansion component to get the output of the second to last laser etc. The polarizing beam splitter transfers the  polarization of $e^{i\omega t}$ and reflects the perpendicular polarization obtained by propagating through a $\lambda/4$ waveplate twice. Alternatively, one could use a Faraday rotator which is non-reciprocal}
\end{figure}

 \section{Parameters of the laser setup}
Here we specify the laser parameters in our Meep simulation \cite{ge2010steady,cerjan2014steady,cerjan2020modeling}
\begin{eqnarray}
n_{1}=\sqrt{\epsilon_{1}+\frac{\gamma_{\perp}\frac{d_{0}}{1+\Gamma\left|E\right|^{2}}}{\omega_{\sigma}-\omega_{a}+i\gamma_{\perp}}},\,\,\,\Gamma=\frac{\gamma_{\perp}^{2}}{\left(\omega_{\sigma}-\omega_{a}\right)^{2}+\left(\gamma_{\perp}\right)^{2}},\,\,\,\,\,\,\,\,\,\,\,\,\,\,\,\,\,\,\,\,\,\,\,\,\,\,\nonumber\\
\gamma_{\perp}=4,\,\,d_{0}=\frac{\theta^{2}}{\hbar\gamma_{\perp}}\left(\frac{\gamma_{12}-\gamma_{21}}{\gamma_{12}+\gamma_{21}}N_{0}\right)=0.06,\,\,\epsilon_{1}=2.25,\,\,\,\,\,\,\,\,\,\,\,\,\,\,\,\,\,\,\,\,\,\,\,\,\,\,\,\,\nonumber\\
\omega_{\sigma}=40.77,\,\,\omega_{a}=40,\,\,l_{1}=1,\,\,n_{1}=1.50298-0.019i,\,\,\,\,\,\,\,\,\,\,\,\,\,\,\,\,\,\,\,\,\,\,\,\,\,\,\,\,\nonumber\\
\,\,\theta=1,\,\,N_0=37,\gamma_{12}=0.005065, \gamma_{21}=0.005,\,\,\,\,\,\,\,\,\,\,\,\,\,\,\,\,\,\,\,\,\,\,\,\,\,\,\,\,\,\,\,\,\,\,\,\,\,\,\nonumber
\end{eqnarray}
where $\gamma_{\perp}$ is the gain bandwidth, $d_{0}$ is the pump amplitude,  $\theta$ is the light-atom interaction strength coefficient, $\gamma_{12}$ is the pump rate, $\gamma_{21}$ is the nonradiative decay rate, $N_0$ is the population density,
$\omega_{a}$ is the atom transition  frequency in the gain medium, $\omega_{\sigma}$
is the lasing frequency, and we have used SALT units \cite{ge2010steady}. Note that our gain value is 1705 1/cm, which is on the order of an experimentally-realized gain at room temperature of 940 1/cm and one can increase $r_1$ to reduce the gain e.g., by introducing Bragg mirrors \cite{hill2014advances}.

\section{Approximating the spontaneous emission in the linear response regime}
Approximately only the resonant mode is dominant in this calculation and we write:
$G_{\mu\mu}=\frac{E_{\mu\mu}}{\omega^{2}p},\mathbf{E}=\mathbf{E}_{0}-\int\frac{kdkd\phi}{\left(2\pi\right)^{2}}\frac{s_{k}}{s-s_{k}}\frac{4\pi p}{\epsilon_{2}}\frac{\mathbf{E}_{k}\left(\mathbf{z}_{0}\right)\mathbf{E}_{-k}\left(\mathbf{z}_{0}\right)}{\left\langle \mathbf{E}_{k}|\mathbf{E}_{k}\right\rangle },$
where $\mathbf{E}_k$ is a full electrodynamic eigenfunction, $s_k=\epsilon_2/(\epsilon_2-\epsilon_{1k}),s=\epsilon_2/(\epsilon_2-\epsilon_{1}),$ $\epsilon_1$ and $\epsilon_2$ are the cavity and free space permittivities, respectively, and $\epsilon_{1k}$ is the eigenpermittivity. $\left\langle \mathbf{E}_{\mathbf{\mathrm{k}}}|\mathbf{E}_{\mathbf{\mathrm{k}}}\right\rangle =\int dV\mathbf{E}_{-\mathrm{\mathbf{k}}}\left(\mathbf{r}\right)\cdot\mathbf{E}_{\mathbf{\mathrm{k}}}\left(\mathbf{r}\right) $ is the inner product in the cavity volume, $z_0$ is the dipole location, and $\mathbf{E}_0$ is the dipole field in a uniform medium with the cavity permittivity \cite{farhi2016electromagnetic,bergman2018spectral,bergman2020scattering}. Note that both the denominator and numerator vanish for $k=0,$ which is the magnitude of the $\mathbf{k}$ vector that is parallel to the interface.

 \section{Excitation of the laser with additional incoming field amplitudes}
 Here we excite the laser with 10 times smaller and 5 times larger values of $E_\mathrm{inc}$ and plot the responses in Fig. 4 (a) and (b). It can be seen that at low values of $E_\mathrm{inc}$ the scattered field becomes wavy after the overshoot and that at large values of $E_\mathrm{inc}$ the response is more rapid.
\begin{figure*}[htp]
\subfloat[]{
\includegraphics[width=8cm]{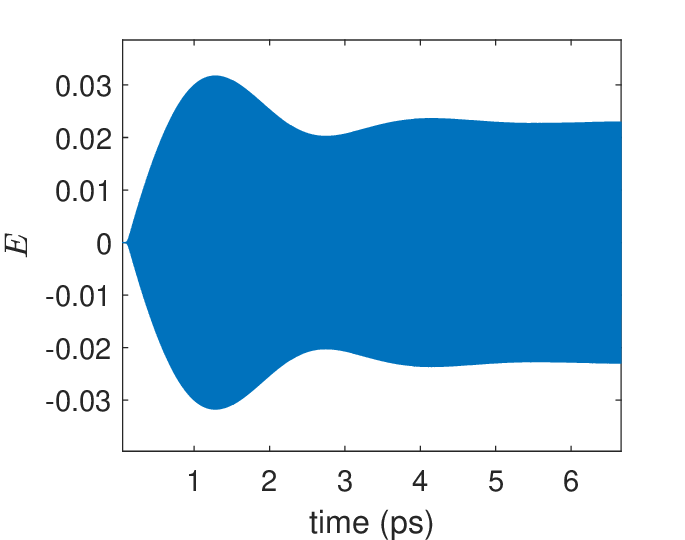}
}\subfloat[]{
\includegraphics[width=8cm]{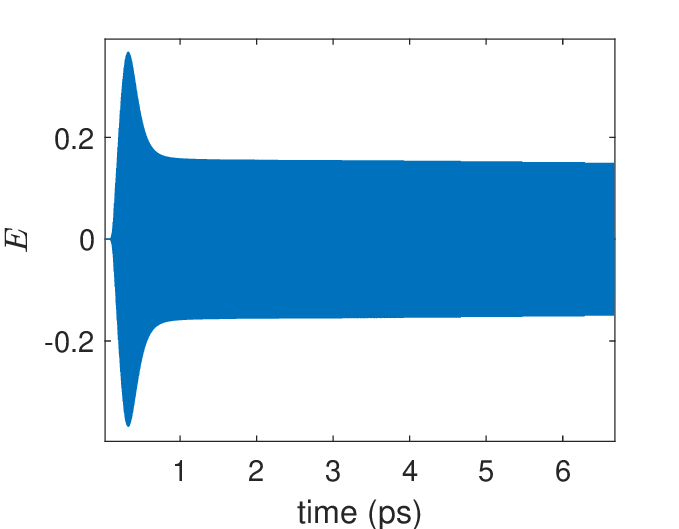}
}

\caption{The laser response for (a) $J_{\mathrm{ext}}=0.001$ and (b) $J_{\mathrm{ext}}=0.05.$}
\end{figure*}

\section{Rate equations for a laser with a constant drive}
We consider the rate equations for a three-level atom with a fast $1\rightarrow 0$ transition (for the I-SALT stability analysis see Appendix in Ref.  \cite{cerjan2014steady}). We can thus use an effective two level description \cite{haken1986laser}.
We write the equations for the photon number $n$ and population at the third level $N_{2}$ with a constant drive
\begin{equation}
\frac{dn}{dt}=-2\kappa n+DWn+T_{1}\frac{E_{\mathrm{inc}}^{2}}{\hbar},
\end{equation}
\begin{equation}
\frac{dN_{2}}{dt}=\left(N-N_{2}\right)w_{20}-w_{12}N_{2}-WN_{2}n,
\end{equation}
where $2\kappa$ is the inverse time of a photon in the cavity, which
for a laser at threshold reads $2\kappa=\frac{cG}{n}$ where $G$
is the gain, $T_{1}$ the transmission coefficient, $w_{20}$ the
pump strength, $w_{12}$ the non-radiative decay coefficient, and $W$
is the stimulated emission coefficient defined as $W=\frac{1}{\tau}\frac{c^{3}}{V8\pi\omega^{2}\Delta\omega}$,
where $\tau$ is the atom lifetime, $V$ is the laser volume, and
$\Delta\omega$ is the atom linewidth.

Since we assume a fast $1\rightarrow 0$ transition, $D=N_{2}-N_{1}\approx N_{2}$ and we write
\begin{equation}
\frac{dD}{dt}=\left(N-D\right)w_{20}-w_{12}D-WDn.
\end{equation}

The stationary solutions read 
\begin{equation}
0=-2\kappa n+DWn+T_{1}\frac{E_{\mathrm{inc}}^{2}}{\hbar},
\end{equation}
\begin{equation}
0=\left(N-D\right)w_{20}-w_{12}D-WDn.
\end{equation}
We substitute $n=n_{0}+\delta n,\,\,\,D=D_{0}+\delta D$ in the equations
and omit the $\delta D\delta n$ terms to get
\[
\frac{d\delta n}{dt}=-2\kappa\left(n_{0}+\delta n\right)+\left(D_{0}n_{0}+n_{0}\delta D+D_{0}\delta n\right)W+T_{1}\frac{E_{\mathrm{inc}}^{2}}{\hbar},
\]
\[
\frac{d\delta D}{dt}=\left(N-\left(D_{0}+\delta D\right)\right)w_{20}-w_{12}\left(D_{0}+\delta D\right)-W\left(D_{0}n_{0}+n_{0}\delta D+D_{0}\delta n\right).
\]

Subtracting the stationary equations we obtain
\[
\frac{d\delta n}{dt}=\left(D_{0}W-2\kappa\right)\delta n+n_{0}W\delta D,
\]
\[
\frac{d\delta D}{dt}=\delta D\left(-w_{20}-w_{12}-Wn_{0}\right)-WD_{0}\delta n.
\]
We then simplify these equations using the stationary equations and get
\begin{equation}
\frac{d\delta n}{dt}=-T_{1}\frac{E_{\mathrm{inc}}^{2}}{\hbar}\frac{1}{n_{0}}\delta n+n_{0}W\delta D,
\end{equation}
\begin{equation}
\frac{d\delta D}{dt}=-\frac{Nw_{20}}{D_{0}}\delta D-WD_{0}\delta n,
\end{equation}
where $D_{0},n_{0}$ can be expressed from the stationary equations.

We guess solutions of the type
\[
\delta n=Ae^{\alpha t},\,\,\,\delta D=Be^{\alpha t},
\]
and obtain
\begin{equation}
\left(\alpha+T_{1}\frac{E_{\mathrm{inc}}^{2}}{\hbar}\frac{1}{n_{0}}\right)A-Bn_{0}W=0,
\end{equation}
\begin{equation}
WD_{0}A+\left(\alpha+\frac{Nw_{20}}{D_{0}}\right)B=0,
\end{equation}
We write for $\alpha$
\[
\left(\alpha+T_{1}\frac{E_{\mathrm{inc}}^{2}}{\hbar}\frac{1}{n_{0}}\right)\left(\alpha+\frac{Nw_{20}}{D_{0}}\right)+n_{0}W^{2}D_{0}=0,
\]
\[
\alpha^{2}+\left(T_{1}\frac{E_{\mathrm{inc}}^{2}}{\hbar}\frac{1}{n_{0}}+\frac{Nw_{20}}{D_{0}}\right)\alpha+T_{1}\frac{E_{\mathrm{inc}}^{2}}{\hbar}\frac{1}{n_{0}}\frac{Nw_{20}}{D_{0}}+n_{0}W^{2}D_{0}=0,
\]
\[
\alpha_{1,2}=-\left(T_{1}\frac{E_{\mathrm{inc}}^{2}}{\hbar}\frac{1}{n_{0}}+\frac{Nw_{20}}{D_{0}}\right)\pm\sqrt{\left(T_{1}\frac{E_{\mathrm{inc}}^{2}}{\hbar}\frac{1}{n_{0}}+\frac{Nw_{20}}{D_{0}}\right)^{2}-4\left(T_{1}\frac{E_{\mathrm{inc}}^{2}}{\hbar}\frac{1}{n_{0}}\frac{Nw_{20}}{D_{0}}+n_{0}W^{2}D_{0}\right)},
\]
\[
\alpha_{1,2}=-\left(T_{1}\frac{E_{\mathrm{inc}}^{2}}{\hbar}\frac{1}{n_{0}}+\frac{Nw_{20}}{D_{0}}\right)\pm\sqrt{\left(T_{1}\frac{E_{\mathrm{inc}}^{2}}{\hbar}\frac{1}{n_{0}}-\frac{Nw_{20}}{D_{0}}\right)^{2}-4n_{0}W^{2}D_{0}},
\]
substituting $2\kappa n_{0}-T_{1}\frac{E_{\mathrm{inc}}^{2}}{\hbar}=D_{0}Wn_{0},$
we get
\[
\alpha_{1,2}=-\left(T_{1}\frac{E_{\mathrm{inc}}^{2}}{\hbar}\frac{1}{n_{0}}+\frac{Nw_{20}}{D_{0}}\right)\pm\sqrt{\left(T_{1}\frac{E_{\mathrm{inc}}^{2}}{\hbar}\frac{1}{n_{0}}-\frac{Nw_{20}}{D_{0}}\right)^{2}-4W\left(2\kappa n_{0}-T_{1}\frac{E_{\mathrm{inc}}^{2}}{\hbar}\right)}.
\]
It is clear that the first term (outside of the square root) increases
when we increase $E_{\mathrm{inc}}$ since $E_{\mathrm{inc}}^{2}$
increases more rapidly than $n_{0}$ (see Fig. 2d), and $D_{0}$ decreases, which
implies fast dynamics. Similar arguments follow for the second term
in the square root that can be negative, which means that at large
values of $E_{\mathrm{inc}}$ there are no oscillations, in agreement
with our simulation results.

\[
n_{0}^{\pm}=\frac{E^{2}T_{1}W-2h\kappa w_{12}-2h\kappa w_{20}+hNWw_{20}\pm\sqrt{(E^{2}T_{1}W+2h\kappa (w_{12}+w_{20})+hNWw_{20})^{2}-8h^{2}\kappa  NW\text{\ensuremath{w_{20}}}(w_{12}+w_{20})}}{4h\kappa  W}
\]
\[
D_{0}^{\pm}=\frac{E^{2}T_{1}W+2h\kappa w_{12}+2h\kappa w_{20}+hNWw_{20}\mp\sqrt{(E^{2}T_{1}W+2h\kappa (w_{12}+w_{20})+hNWw_{20})^{2}-8h^{2}\kappa 
 NW\text{\ensuremath{w_{20}}}(w_{12}+w_{20})}}{2hW(w_{12}+w_{20})}
\]
Under the assumption of a small drive we get
\[
\sqrt{(E^{2}T_{1}W+2h\kappa (w_{12}+w_{20})+hNWw_{20})^{2}-8h^{2}\kappa NW\text{\ensuremath{w_{20}}}(w_{12}+w_{20})}=
\]
\[
\sqrt{\left(E^{2}T_{1}W\right)^{2}+2E^{2}T_{1}W\left(2h\kappa (w_{12}+w_{20})+hNWw_{20}\right)+(2h\kappa (w_{12}+w_{20})+hNWw_{20})^{2}-8h^{2}\kappa NW\text{\ensuremath{w_{20}}}(w_{12}+w_{20})},
\]
\[
a\equiv (2h\kappa (w_{12}+w_{20})+hNWw_{20})^{2}-8h^{2}\kappa NW\text{\ensuremath{w_{20}}}(w_{12}+w_{20}),
\]
\[
\sqrt{a+\left(E^{2}T_{1}W\right)^{2}+2E^{2}T_{1}W\left(2h\kappa (w_{12}+w_{20})+hNWw_{20}\right)}=
\]
\[\sqrt{a\left(1+\frac{\left(E^{2}T_{1}W\right)^{2}+2E^{2}T_{1}W\left(2h\kappa (w_{12}+w_{20})+hNWw_{20}\right)}{a}\right)}\]
\[
\approx\sqrt{a}\left(1+\frac{\left(E^{2}T_{1}W\right)^{2}+2E^{2}T_{1}W\left(2h\kappa (w_{12}+w_{20})+hNWw_{20}\right)}{a}\right).
\]
we get (choosing the physical solutions)
\[
n_{0}^{+}\approx\frac{E^{2}T_{1}W-2h\kappa w_{12}-2h\kappa w_{20}+hNWw_{20}+\sqrt{a}\left(1+\frac{\left(E^{2}T_{1}W\right)^{2}+2E^{2}T_{1}W\left(2h\kappa (w_{12}+w_{20})+hNWw_{20}\right)}{2a}\right)}{4h\kappa W},
\]
\[
D_{0}^{+}\approx\frac{E^{2}T_{1}W+2h\kappa w_{12}+2h\kappa w_{20}+hNWw_{20}-\sqrt{a}\left(1+\frac{\left(E^{2}T_{1}W\right)^{2}+2E^{2}T_{1}W\left(2h\kappa (w_{12}+w_{20})+hNWw_{20}\right)}{2a}\right)}{2hW(w_{12}+w_{20})}.
\]
Since
$\frac{\left(2h\kappa(w_{12}+w_{20})+hNWw_{20}\right)}{\sqrt{a}}>1,$
$n_{0}^{+}$ and $D_{0}^{+}$ increase and decrease when increasing $E^{2},$ respectively.

We can express  $E_\mathrm{sat}$ with $n_0$ 
$$\frac{dn}{dt}=-2n_{0}\kappa=-P/\hbar=-E_{\mathrm{sat}}^{2}/\hbar,$$
$$E_{\mathrm{sat}}=\sqrt{2\hbar n_{0}\kappa}=\sqrt{\frac{\hbar n_{0}}{2nl_1}\ln\left(R_{1}\right)}=\sqrt{\frac{\hbar n_{0}}{2nl_1}\ln\left(R_{1}\right)}.$$

Finally, we phenomenologically approximate $W$
$\frac{dn}{dt}=N_{2}Wn,\,$
$n=Ae^{N_{2}Wt},\,\,
E=e^{kn_{i}\Delta l}=e^{kn_{i}c/n_r\Delta t},\,\,
E^{2}=e^{2kn_{i}c/n_r\Delta t},\,\,
N_{2}W=2kn_{i}c/n_r,\,\,
W\approx 2kn_{i}c/\left(n_r N_{2i}\right).$



%
%
%
%
%
\section{The range of $\Gamma_z\mathrm{s}$ that can be generated}
We now evaluate the modulation frequency range that can be used in this platform. First, it can be readily seen that for the waveforms within this class there are no observable deviations from the analytic expression (except for a short transient) since there is no dicretization of the signal and we conclude that their modulation frequency is very high. Then, for expanded waveforms, we note that for a single Taylor expansion, we require $\Gamma t_{f}\lesssim3,$ where $t_{f}$ is the final pulse time, for the expansion to be accurate using five waveforms. Then, for high-modulation frequencies since the output is assumed to be valid from a certain time, we also require that $t_{f}\gtrsim\max \left(2ln/c,1/\Delta\omega\right),$ where $\Delta\omega$ is a width in which the single-pole approximation is valid and is on the order of the free spectral range and also depends on the laser linewidth and  gain bandwidth. Note that for long cavities one can require $t_{f}\lesssim2ln/c,\,\,t_{f}\gtrsim1/\Delta\omega$ but $\Delta\omega$ becomes small. Thus, we get $\Gamma\lesssim3/\max\left(2ln/c,1/\Delta\omega\right)$ that we approximate to be on the order of $\omega/5$ for our laser. This limitation may be overcome by utilizing an ENZ laser \cite{jia2021broadband} or reexpanding the output. On the other hand, for the Taylor expansion to capture significant temporal variations, one can require $\Gamma t_{f}\gtrsim1,$ and due to the gain saturation we impose $t_{f}<t_{\mathrm{sat}}.$ Thus, we get $\Gamma\gtrsim1/t_{\mathrm{sat}},$ which in our case is 2THz. This limit may also be overcome by using a long ENZ laser. In conclusion, we obtain $1/t_{\mathrm{sat}}\lesssim\Gamma\lesssim3/ \max \left(2ln/c,1/\Delta\omega\right).$

 \end{widetext}
\end{document}